\documentclass[journal]{IEEEtran}
\usepackage{amsmath}
\usepackage{amssymb}
\usepackage{amsfonts}
\usepackage{graphicx}

\usepackage{geometry}                
\usepackage{graphicx}
\usepackage{amssymb}
\usepackage{epstopdf}
\usepackage{amsfonts}
\usepackage{amsthm}
\usepackage{verbatim}
\usepackage{amsmath, amssymb}
\usepackage{tikz}
\usetikzlibrary{matrix, arrows}
\usepackage{listings}
\usepackage{color}
\usepackage{listings}
\usepackage[all]{xy}
\usepackage[pdftex,colorlinks,linkcolor=blue,citecolor=blue]{hyperref}
\usepackage{graphicx}
\usepackage{float}

\newcommand{\bC}{\mathbb{C}}
\newcommand{\C}{\mathbb{C}}

\newcommand{\bN}{\mathbb{N}}

\newcommand{\bR}{\mathbb{R}}
\newcommand{\R}{\mathbb{R}}

\newcommand{\bZ}{\mathbb{Z}}
\newcommand{\eps}{\varepsilon}

\newcommand{\E}{\mathbb{E}}

\usepackage{dsfont}
\newcommand{\indic}{\mathds{1}}

\newcommand{\Norm}{\mathcal{N}}
\newcommand{\cvdis}{\Longrightarrow}

\theoremstyle{definition}
\newtheorem*{myproof}{Proof}

\newcommand{\Prob}{\mathbb{P}}

\usepackage{enumitem}

\newcounter{dummy} \numberwithin{dummy}{section}

\theoremstyle{definition}
\newtheorem{mydef}[dummy]{Definition}

\newtheorem{thm}[dummy]{Theorem}
\newtheorem{lemma}[dummy]{Lemma}

\newtheorem{remark}[dummy]{Remark}

\newtheorem*{thm*}{Theorem}
\newtheorem*{notation*}{Notation}

\begin{document}
\title{Delocalisation of one-dimensional marginals of product measures and the capacity of LTI discrete channels}
\author{Maxime~Bombar~\IEEEmembership{}
    and~Alexander~Fish~\IEEEmembership{}
\thanks{M. Bombar is with ENS Paris-Saclay, Cachan, France, e-mail: maxime.bombar@ens-paris-saclay.fr.}
\thanks{A. Fish is with School of Mathematics and Statistics, University of Sydney, Sydney, NSW 2000, Australia,
e-mail: alexander.fish@sydney.edu.au.}}
\maketitle
\begin{abstract}
We consider discrete linear time invariant (LTI) channels satisfying the phase independence (PI) assumption. We show that under the PI assumption the capacity of LTI channels is positive. The main technical tool that we use to establish the positivity of the capacity is the delocalisation theorem for one-dimensional marginals of the product measure due to Ball and Nazarov. We also prove two delocalisation results that can be seen as extensions of Ball-Nazarov Theorem.
\end{abstract}
\begin{IEEEkeywords}
LTI channels, PI assumption, Shannon capacity, delocalisation.
\end{IEEEkeywords}
\section{\textbf{Introduction}}
\label{Section1}
In this work we consider linear time-invariant (LTI) discrete channels.
 Given a sequence $\mathbf{x} \in \bC^N$ at the transmitter side, we assume that at the receiver we obtain the sequence $\mathbf{y} \in \mathbb{C}^N$ given by:
\begin{equation}
\label{equation1}
\mathbf{y}[n] = \sum_{m=0}^{N-1} a_{m} \mathbf{x}[n-m \mbox{ mod } N] + \mathbf{\omega}[n], 
\end{equation}
\[
n=1,\ldots,N,
\]
where $\{a_m\}$'s are the attenuation coefficients, and $\mathbf{\omega}$ is the additive white Gaussian noise (AWGN).
The parameter $N$ will be the total length of the  sequence. We will see in Section \ref{Section2} that in the case of wireless communication we have $N = T W$, where $W$ is the bandwidth and $T$ is the time of the modulation. We will assume that the signal to noise ratio (SNR) is fixed and equal to $P$.
Let us denote by $\alpha_m = | a_m |$, and $X_m = arg(a_m)$, $m = 0,\ldots, N-1$. Then we rewrite equation (\ref{equation1}) by use of the operator notation:
\begin{equation}
\label{eq2}
\mathbf{y} = H_N(\mathbf{x}) + \mathbf{\omega}.
\end{equation}
The operator $H_N:\mathbb{C}^N \to \mathbb{C}^N$ is given by:
\begin{equation}
\label{channel_eq}
H_N = \sum_{m=0}^{N-1} \alpha_m X_m \pi_m,
\end{equation}
where $\pi_m$ is the circular shift by $m$ on $\mathbb{C}^N$, i.e., $\pi_m(x)[n] = x[n-m \mbox{ mod } N]$, $n = 0,\ldots,N-1$.
\smallskip

We introduce the following assumption for requirements on the channel operator.
\smallskip

\begin{mydef}[PI assumption]
We will say that the channel operator $H_N$ given by Equation (\ref{channel_eq}) satisfies the phase independence (PI) assumption, if all $X_m$'s are independent identically distributed random variables having uniform distribution on the unit circle.
\end{mydef}
\smallskip

\begin{remark}
In the case of a channel appearing in wireless communication under assumption that the carrier frequency $f_c$ is much higher than the bandwidth $W$, it is natural to assume that $arg(a_m)$'s are independent identically distributed random variables having uniform distribution on the unit circle (see Section \ref{Section2}).
\end{remark}

We further assume that at the receiver the channel estimation is performed, and we know the attenuation coefficients $\{a_m\}$'s. We will say that an $N$-dimensional probability distribution $\mathbf{x}$ satisfies $A_{PN}$, if $\E(tr(\mathbf{x} \mathbf{x}^t)) = PN$. We will denote the latter by $\mathbf{x} \in A_{PN}$. The main result of the paper is the following.

\begin{thm}
\label{main-thm}
The asymptotic Shannon capacity of the channel given by (\ref{eq2}) and (\ref{channel_eq}), and satisfying the PI assumption is positive:
\[
\liminf_{N \to \infty} \frac{1}{N} \sup_{\mathbf{x} \in A_{PN}}I(\mathbf{x}; (\mathbf{y},H_N)) > 0.
\]
Here $I(\mathbf{x},\mathbf{y})$ denotes the mutual information of the vectors $\mathbf{x}$ and $\mathbf{y}$, see \cite{S}.
\end{thm}
\smallskip

In order to prove Theorem \ref{main-thm} we use the probabilistic result of Ball and Nazarov \cite{BN} on uniform delocalisation of  one-dimensional marginals of product measure for a given delocalised probability distribution in $\bR^d$. Let us recall that for a $\bR^d$-valued random variable $X$, the Levy concentration function of $X$ is defined for every $\eps > 0$ by
\[
C_{\eps}(X) = \sup_{z \in \bR^d} \Prob(\| X - z \| \le \eps).
\]
To state Ball-Nazarov theorem, we will also have to define the class $\mathcal{F}$ of the random variables:
\begin{align*}
\mathcal{F} = \{ \sum_{n=1}^{N} a_n X_n \,\, | & \,\,N \ge 1, \,\, \sum_{n=1}^N |a_n|^2 = 1, \\
 & X_1,\ldots,X_N \mbox{ are i.i.d. as } X \}.
\end{align*}

Then the following theorem holds true:
\smallskip

\begin{thm}[Ball-Nazarov \cite{BN}]
\label{thm2}
Let $X$ be a $\bR^d$-valued random variable. Then there exists a universal constant $A > 0$ such that for every $\eps > 0$ we have
\[
\sup_{ S \in \mathcal{F}} C_{\eps}(S) \le A C_{\eps}(X).
\]
\end{thm}
\smallskip
\begin{remark}
The quantitative bound $A=\sqrt{2}$ in the case where $X$ has a bounded density in $\mathbb{R}$ has been recently obtained by Rudelson and Vershynin (Theorem 1.2 in \cite{RV}). Moreoever, Theorem of Rudelson-Vershynin does not require from the random variables to be identically distributed, but only independent. Their theorem is a corollary of K. Ball Theorem, see \cite{B}, on the slice of a maximal area of the unit cube in $\mathbb{R}^N$.
\end{remark}
\smallskip

In this paper, we prove a stronger version of Theorem \ref{thm2} in $\bR^2$ under more restrictive conditions. We will identify $\bR^2$ with the complex plane $\bC$. We assume that $X$ is uniformly distributed on the unit circle in $\mathbb{C}$, and
there exists $\eps_0 > 0$ such that a projection of the unit weight vector on any $4$ coordinates has $\ell_2$-norm at most $1-\eps_0$.
More precisely, let us denote by $\mathcal{S}_{4,N}$ the collection of all $4$-dimensional subspaces in $\bR^N$ generated by any $4$ vectors from the standard basis $\{e_1,\ldots,e_N\}$ of $\bR^N$. We define the class
\begin{align*}
S_{4,\eps_0} = & \\
 \{ (a_1,\ldots,a_N) \in S^{N-1} \, | &\, \| \pi_{E}(a) \|_2^2 \le 1 - \eps_0, \\ &\forall E \in \mathcal{S}_{4,N}, N \ge 5\},
\end{align*}
where $\pi_E$ denotes the orthogonal projection on the subspace $E$, and $S^{N-1}$ is the unit sphere in $\bR^N$ defined by
\[
S^{N-1} = \{(a_1,\ldots,a_{N}) \in \bR^N \, | \, \sum_{k=1}^{N} a_k^2 = 1 \}.
\]
 We think on the set $S_{4,\eps_0}$ as the set of non too sparse power profiles of the channel operator $H_N$.
We define the collection of $\bC$-valued random variables $\mathcal{F}_{4,\eps_0}$ to be:
\[
\mathcal{F}_{4,\eps_0} =
\]
\[
 \{ a_1 X_1 + \ldots + a_N X_N \in \mathcal{F} \, | \, (a_1,\ldots,a_N) \in S_{4,\eps_0}\}.
\]
We prove the following result\footnote{Our method can be also used to provide an alternative proof of Theorem  \ref{thm2} in the special case, where $X$ is uniformly distributed on the unit circle in $\bC$.} which is slightly better than the bound that we can deduce from Theorem \ref{thm2}.
\smallskip

\begin{thm}
\label{thm3}
Let $X$ be a random variable uniformly distributed on the unit circle in $\bC$, and $\eps_0 > 0$. Then there exists a constant $B > 0$ such that for every $\eps > 0$ we have
\[
\sup_{ S \in \mathcal{F}_{4,\eps_0}} C_{\eps}(S) \le B \eps^2.
\]
\end{thm}

Theorem \ref{thm3} is sharp. We know that the $3$-fold convolution of the uniform probability measure on the unit circle has density function $f(x)$ in the disk of radius $3$ in $\bC$ with singularities on the unit circle. More precisely, there are constants $C,c > 0$ such that for $x \in \bC$ close to the unit circle we have \cite{CFOT}:
\[
c \left| \log(1 - |x|) \right| \le f(x) \le C \left| \log(1 - |x|) \right|.
\]
The latter implies that if $S = \frac{1}{2}(X_1+X_2+X_3+X_4)$, where $X_1,\ldots,X_4$ are independent uniformly distributed on the unit circle, then there exists $K > 0$ such that
\[
\Prob(|S| \le \eps) \ge K \eps.
\]

Another consequence of our approach to the delocalisation is a multi-dimensional analog of Rudelson-Vershynin's result (Theorem 1.2 in \cite{RV}).

\begin{thm}
\label{mult-dim-deloc-thm}
Let $K, L, M, \delta, \eta > 0$. Let $(X_n)$ be $\bR^d$-valued independent random variables, having densities bounded  by $K$, and $\Gamma_n$ be the covariance matrices\footnote{Let $\mu_n := \E(X_n)$ be the expection of $X_n$. Then $\Gamma_n := \E((X_n - \mu_n)(X_n - \mu_n)^t)$ is the covariance matrix of $X_n$. } of $X_n$'s.
Assume that for all $n$, all the entries of $\Gamma_n$'s are bounded by $L$, $\det \Gamma_n \ge \delta$, and $\E[\|X_n\|^{2+\eta}] \le M$. Then there exists $C > 0$ such that for every $N \ge 1$, the density function of $a_1X_1 + \ldots + a_N X_N$ is bounded by $CK$, provided that $\sum_{n=1}^{N} a_n^2 = 1$.
\end{thm}

%
%
%
%

 \section{\textbf{Linear Time Invariant (LTI) discrete channels}}
 \label{Section2}

 In this Section we will justify the channel model given by the Equations (\ref{eq2}) and (\ref{channel_eq}) in the scenario of wireless communication in urban environment. Our assumptions are the following:
 \bigskip

 \begin{itemize}
 \item The communication is done between a transmitter and a receiver by use the frequency band $\left[f_c - \frac{W}{2}, f_c + \frac{W}{2}\right]$, where $f_c$ is the carrier frequency and $W$ is the bandwidth. Also, we assume $f_c \gg W$.
  \vspace{\baselineskip}
 \item The environment, the transmitter and the receiver are almost static (have very low speed). As a result the signal experiences a negligible Doppler effect. The environment has many obstacles, and there is no line of sight between the transmitter and the receiver.
  \vspace{\baselineskip}
 \item The processing is performed at the baseband $\left[- \frac{W}{2}, \frac{W}{2}\right]$.
 \end{itemize}
\smallskip

 Let us assume that at the transmitter side we have a sequence $S[n] = b_n, \quad n = 1,2,\ldots,$ of real (complex) numbers that we would like to send to a receiver. It is a standard procedure in wireless communication, to use some Digital-to-Analog transform, in order to generate a function $s(t)$ with a spectral profile in a given band which encompasses the information. We will stick in this paper to the Shannon-Nyquist transform given by
\[
s(t) = exp(i f_c t)\left(\sum_{m} b_m sinc(-Wt + m)\right),
\]
where $sinc(t) = \frac{\sin{\pi t}}{\pi t}$. It follows from the properties of the sinc function that:

\begin{itemize}
\item The signal $s(t)$ satisfies that $\widehat{s} \subset \left[f_c - \frac{W}{2}, f_c + \frac{W}{2}\right]$.\\
\item The Inverse (Analog-to-Digital) operator is given by the sampling $s(t)$ at the time slots $\{\frac{1}{W},\frac{2}{W},\ldots\}$. In other words, we have
\[
s\left(\frac{n}{W}\right) = b_n = S[n], \mbox{ for } n \ge 1.
\]
\end{itemize}

We use the model of multi-path propagation of a signal \cite{TV}. Assume that at the transmitter side we generate the signal $s(t)$ as above. Then for any path $\mathcal{L}$ of length $\ell$ we obtain at the receiver the signal:
\[
r_{\mathcal{L}}(t) = \alpha_{\mathcal{L}} s(t-\ell/c) + \omega_{\mathcal{L}}(t),
\]
where $c$ denotes the speed of light, $\alpha_{\mathcal{L}} \in \bR$ is the attenuation coefficient which depends on the path length (decays as $\frac{1}{\ell^2}$) and on the environment, and $\omega_{\mathcal{L}}$ denotes the additive noise along the path $\mathcal{L}$.

We assume that the received signal $r_{\mathcal{L}}(t)$ is shifted to the baseband first, and then, we sample \\
$exp(-i f_c t) r_{\mathcal{L}}(t)$ at the time slots $\{\frac{1}{W},\frac{2}{W},\ldots\}$. We also assume, as in \cite{FGHSS}, that the time shift $\frac{\ell}{c}$ lies on the lattice $\frac{1}{W} \bZ$, i.e., there exists $n_{\mathcal{L}} \in \bZ$ such that $\ell/c = \frac{n_{\mathcal{L}}}{W}$. Then, at the receiver side, we obtain the following sequence:

\begin{align*}
 R_{\mathcal{L}}[n] & = exp(-i f_c n_{\mathcal{L}}/W) r_{\mathcal{L}}\left( \frac{n}{W} \right) \\
           & = exp(-i f_c n_{\mathcal{L}}/W) \alpha_{\mathcal{L}} b_{n - n_{\mathcal{L}}} + \omega_{\mathcal{L}}(n/W),\\
           & \quad n=1,2,\ldots
\end{align*}

Therefore, we have
\begin{align*}
 R_{\mathcal{L}}[n] = exp(-i f_c n_{\mathcal{L}}/W) \alpha_{\mathcal{L}} & S[n-n_{\mathcal{L}}] \\
                                      & + \omega_{\mathcal{L}}(n/W).
\end{align*}
By our assumptions, $f_c \gg W$, and therefore, the phase $f_c n_{\mathcal{L}}/W$ will change drastically if we replace $n_{\mathcal{L}}$ by another close to it integer. So, we approximate $R_{\mathcal{L}}$ by the following:
\[
R_{\mathcal{L}}[n] \approx \alpha_{\mathcal{L}} X_{\mathcal{L}} S[n-n_{\mathcal{L}}] + \omega_{\mathcal{L}}(n/W),
\]
where $X_{\mathcal{L}}$ is uniformly distributed random variable on the unit circle in $\bC$.
\smallskip

Finally, taking all the paths from the transmitter to the receiver, by the law of superposition, we obtain the sequence $R$ given by:
\[
R[n] = \sum_{k \in \bN} \alpha_{k} X_k S[n-k] + \omega[n], \quad n = 1, 2,\ldots
\]
where $\omega[n] = \sum_{\mathcal{L}} \omega_{\mathcal{L}}(n/W)$. We will assume that the phases $X_k$  are independent uniformly distributed on the unit circle in $\bC$, since they belong to different paths, and therefore are not related one to each other. 
 It is a standard assumption that all $\omega[n]$'s are independent complex Gaussian random variables with zero mean and variance one.
\smallskip

By use of the trick of periodic prefix, like in \cite{FGHSS}, using a transmission of a finite duration time $T$, we can assume that the transmitted and received sequences, $S$ and $R$ respectively, have length $N = T W$, and the relation between $S$ and $R$ is described by equation:

\begin{equation}
\label{eq-channel}
R[n] = \sum_{k=0}^{N-1} \alpha_{k} X_k S[n-k \mbox{ } mod \mbox{ } N] + \omega[n],
\end{equation}
where $n = 0,1, \ldots, N-1$. Notice, that the Equation (\ref{eq-channel}) coincides with the model Equation (\ref{channel_eq}) of the channel given in Section \ref{Section1}.

\section{\textbf{On the capacity of LTI channels under the PI assumption}}
\label{section-proof-main-thm}
\smallskip

For each $N$, let $H_N$ be the channel operator defined by Equation (\ref{channel_eq}). It is a standard fact that $H_N$ is diagonalisable and its eigenvectors are the exponential functions $e^{(N)}_k \in \bC^N$ given by
\[
 e^{(N)}_k[m] = e^{\frac{2\pi i k m}{N}}, \quad m \in \{0,1,\ldots,N-1\}.
\]
Therefore,
the matrix of the operator $H_N$ in the basis of the exponentials is  $\left[\begin{array}{ccc} \lambda^{(N)}_1 & \dots & 0 \\ & \ddots & \\ 0 & \dots & \lambda^{(N)}_N\end{array}\right]$ where the $\lambda^{(N)}_k$ are the eigenvalues of $H_N$ given by:

\begin{equation}
 \label{eqlambda}
 \lambda^{(N)}_k = \sum_{m=0}^{N-1}\alpha_mX_me^{-\frac{2\pi i k m}{N}}, \quad k = 1,2,\ldots,N.
\end{equation}

\begin{remark}
 \label{identical}
 Since the uniform distribution on the circle is invariant by rotations, $(\lambda^{(N)}_k)$ is a sequence of identically distributed random variables (but \textbf{not} independent) for each $N$.
\end{remark}

We express $\omega := \omega_1 e_1 + \dots + \omega_N e_N$, where $e_1,\ldots,e_N$ is the basis of the exponentials. Since $\omega$ is a white Gaussian noise of total power $N$, the $\omega_k$ are distributed according to the complex symmetric Gaussian distribution, with 0 mean and variance $1$. We assume that the noise affects each coordinate independently, ie $\E[\omega\omega^t] = I_N$.

\subsection{\textbf{Proof of Theorem \ref{main-thm}}}\
\smallskip

We denote by $\displaystyle C_N = \sup_{\mathbf{x}\in A_{PN} }I(\mathbf{x}; (\mathbf{y}, H_N))$. In this Section we prove the lower bound for $C_N$ which implies Theorem \ref{main-thm}.
\smallskip

In \cite{T}, E. Telatar studied AWGN linear channels\footnote{the noise is assumed to be an additive white Gaussian noise and the operator acts linearly on the signals} with random attenuation coefficient matrix and proved that if $\mathbf{x}$ is constrained to have covariance matrix $Q$ then the choice of $\mathbf{x}$ that maximises $I(\mathbf{x}; (\mathbf{y}, H_N))$, where $H_N$ is random and known at the receiver, is the circularly symmetric complex Gaussian of covariance $Q$ and that

\begin{align*}
& \sup_{\mathbf{x}\in A_{PN}} I(\mathbf{x}; (\mathbf{y}, H_N)) \\
& = \sup_Q \E[\log \det(I_N + H_N Q H_N^{\dag})],
\end{align*}
where the supremum is taken over the choices of non-negative definite $Q$ subject to $tr(Q)\le PN$, and $H_N^{\dag}$ denotes the adjoint operator to $H_N$.

Since for any matrices $A, B$ with suitable sizes $\det(I + AB) = \det(I + BA)$, we can rewrite $C_N$ as follows:

\begin{align}
 C_N & = \sup_Q \E\left[\log \det(I_N + Q H_N^{\dag}H_N)\right] \nonumber \\
   & = \sup_Q \E\left[\log \det(I_N + Q |H_N|^2)\right], \label{information}
\end{align}
where $|H_N|^2$ denotes the diagonal matrix whose coefficients are $|\lambda_i|^2$'s.
Now, taking $Q = PI_N$ we have:

\begin{align*}
 C_N & \ge \E\left[\log \det(I_N + P|H_N|^2)\right] \\
 & = \sum_{m=1}^N \E\left[\log (1 + P|\lambda_m|^2)\right] \\
   & = N \E\left[\log (1 + P|\lambda_1|^2)\right].
\end{align*}

In the last transition we used Remark \ref{identical}.

Then for any $\eps>0$ we have:

\begin{align*}
 \log (1 + P|\lambda_1|^2) & \ge \log (1 + P|\lambda_1|^2)\indic_{\{|\lambda_1| \ge \eps\}} \\
                   & \ge \log (1 + P\eps^2)\indic_{\{|\lambda_1| \ge \eps\}} \\
\end{align*}
The linearity and monotonicity of the expectation imply:
\[
 \E \left[ \log (1 + P|\lambda_1|^2) \right] \ge \log (1 + P\eps^2) \Prob\{|\lambda_1| \ge \eps\}.
\]


By Theorem\footnote{At this point of the proof, we could, alternatively,  use Theorem \ref{thm3} instead of Ball-Nazarov Theorem \ref{thm2}. Our approach would require the delocalisation estimates for $2$, $3,$ and $4$-fold  convolutions of the uniform measure on the unit circle.} \ref{thm2}, there exists\footnote{Indeed, if $X$ is uniformly distributed on the unit circle in $\bC$, then there exists $B > 0$ such that for every $\eps > 0$ we have $C_{\eps}(X) \le B \eps$.} $B > 0$ such that for all $\eps > 0$ we have:
\[
  \Prob\{|\lambda_1| < \eps\} \le B \eps.
\]
Finally, taking $\eps = \dfrac{1}{2 B}$ we obtain:

\[
 \dfrac{1}{N}C_N \ge \dfrac{1}{2}\log\left(1+\dfrac{P}{4B^2}\right) > 0.
\]
Taking the $\liminf$ of the left hand side as $N\to\infty$ concludes the proof of Theorem \ref{main-thm}.


\qed

\section{\textbf{On delocalisation of one-dimensional marginals of the product measure}}
\label{Section3}

\subsection{\textbf{Preliminaries}}

The uniform measure on a circle is not absolutely continuous with respect to Lebesgue measure. However, we can compute the 2-fold convolution of this measure to see that it is absolutely continuous (see \cite{CFOT}). By Lemma \ref{Sum has density}, a linear combination of random variables uniformly distributed on the unit circle also has a density. In  Appendix \ref{Appendix} we will prove the following lemmata:

\begin{lemma}
 \label{Sum has density}
Let $X, Y$ be $\R^d$-valued independent random variables, such that $Y$ has density. Then for every real number $a$, the random variable $aX + Y$ has density. If, moreover, the density of $Y$ is bounded by $M$, then the density of $aX+Y$ is  also bounded by $M$.
\end{lemma}

\begin{remark}
  If both $X$ and $Y$ have a density, then this result directly follows from Young inequality.
\end{remark}

\begin{lemma}
 \label{Bound sum of five}
 Let $a_1, \dots, a_5 \in \C, \; b_1, \dots, b_5>0$ such that for every $k,\, |a_k| \ge b_k$. Let $X_1, \dots, X_5$ be independent random variables uniformly distributed on the unit circle. Then there exists a function $\varphi$ which only depends on the $b_i$'s such that the density $f$ of $a_1 X_1 + \dots + a_5 X_5$ satisfies:
\[
\|f\|_{\infty} \le \varphi(b_1, \dots, b_5).
\]
\end{lemma}


\begin{lemma}
  \label{Bound affine}
  Let $X$ be an $\R^d-$valued random variable having density $f$, $a\in \R \setminus \{0\},\, b\in\R^d$. Then $aX+b$ has density at $x \in \bR^d$ equal to $\dfrac{1}{|a|^d}f\left(\dfrac{x-b}{a}\right)$.
\end{lemma}

\vspace{\baselineskip}

In our work, we use a local $\R^d-$valued Central Limit Theorem that follows from Theorem $1$ in \cite{Sherv}, see Appendix \ref{AppendixB}.

\begin{thm}[Local $\R^d-$valued CLT]
  \label{Local CLT}
  For each $n$, let $X_{n,m}, 1\le m \le n$ be a triangular array of independent $\R^d-$valued, centered, random variables with covariance matrices $C_{n,m} := \E[X_{n,m}X_{n,m}^t]$.
  Let $\sigma^2_{n,m} := tr(C_{n,m}) = \E[\|X_{n,m}\|_2^2]$.\
  
\noindent If
\begin{enumerate}[label=\textbf{\roman*}), align=left]
  \item \label{Convergence}There exists  a symmetric, positive-definite $d\times d$ matrix $C$ such that:
    \[
      \displaystyle \sum_{m=1}^n C_{n,m} \rightarrow C
    \]
  \item \label{Feller}For any $\eps>0$, for any $\theta\in\R^d$,
    \[
      \sum_{m=1}^n \E\left[|\langle \theta, X_{n,m} \rangle|^2\indic_{\{|\langle\theta, X_{n,m}\rangle|>\eps\}}\right] \rightarrow 0
    \]
  \end{enumerate}
   
\noindent then, the sum $S_n := X_{n, 1} + \ldots + X_{n,n}$ converges in distribution towards the $d-$variate Gaussian with $0$ mean and covariance  $C$.

\noindent Moreover, if  one of the following conditions holds true:

  \begin{enumerate}[label=\textbf{\arabic*}), align=left]
  \item Starting from some $n$, $S_n$ has a density, and there exists an integer $p>1$ such that $\|\phi_{n,m}\|_{L_p}^{\frac{2p}{d}}\sigma_{n,m}^2$ is uniformly bounded, where $\phi_{n,m}$ denotes the Fourier transform of $X_{n,m}$.
    \vspace{0.5cm}
  \item All the $X_{n,m}$'s have uniformly bounded density.
  \end{enumerate}
\noindent then

  \[
    \|f_n - g_C\|_{\infty} \rightarrow 0,
  \]
where $f_n$ and $g_C$ denote the densities of $S_n$ and of the $d-$variate Gaussian with $0$ mean and covariance matrix $C$, respectively.

\end{thm}

\subsection{\textbf{Proof of Theorem \ref{thm3}}}

Assume that, contrary to the assertion of Theorem \ref{thm3}, for each $n$ there exist $z \in \bC$, $(a^{(n)})\in S^{(n-1)}$, $\eps_n \rightarrow 0$ such that
\[
  \dfrac{\Prob\{|S_n - z|<\eps_n\}}{\eps_n^2} \rightarrow +\infty,
\]
where $S_n = \sum_{k=1}^n a_k^{(n)}X_k$. In the following, $f_n$ will denote the density function of $S_n$ (for $n\ge 2$) and $g_{\sigma^2}$ will be the density function of a circular symmetric Gaussian with 0 mean and variance $\sigma^2$.
\smallskip
\subsubsection{\textbf{Case 1: Uniform decay of the coefficients}}
\label{Case1}
\smallskip

If we have
\begin{equation}
  \label{Vanish}
  \max_{1\le k\le n}{|a_k^{(n)}|} \rightarrow 0 \mbox{ as } n \to \infty,
\end{equation}
then by Theorem \ref{Local CLT} (see Appendix \ref{AppendixC}),
\[
  \|f_n - g_1\|_{\infty} \rightarrow 0.
\]
Therefore, $\|f_n\|_{\infty} \le \frac{1}{2} + \|g_1\|_{\infty}$ for $n$ sufficiently large. Then
\begin{align*}
  \Prob\{|S_n - z| < \eps_n\} & = \int_{B(z,\eps_n)}f_n(x)dx \\
                              & \le M\eps_n^2
\end{align*}
for some constant $M$ independent of $n$ which leads to a contradiction.
\smallskip
\subsubsection{\textbf{Case 2: Up to \texorpdfstring{$4$}{4} bounded away from zero coefficients}}\
\label{Case2}
\smallskip

There exist $k\in\{1,\dots,4\}$ and $\delta_1,\dots,\delta_k >0$, and there exist a sequence $\varphi(n)$ and $i_1(n) < \dots < i_k(n)$ such that for all $j\in\{1,\dots,k\}$ we have
\[
|a_{i_j(n)}^{(\varphi(n))}| \ge \delta_j,
\]
and 
\[
\max_{\ell \not \in \{i_1(n),\ldots,i_k(n)\}} | a_{\ell}^{(\varphi(n))} | \rightarrow 0 
\]

By the compactness argument, we can assume without loss of generality that:

\[
a_{i_j(n)}^{(\varphi(n))} \substack{\rightarrow \\ k\to\infty} \gamma_{i_j} \ge \delta_j.
\]

By the symmetry, we will assume that $i_j(n) = j$ for every $n$. Since the coefficient vectors are in $\mathcal{F}_{4,\eps_0}$, it follows that
\[
\sum_{i=1}^k \gamma_i^2 \le 1 - \eps_0.
\]
By Theorem \ref{Local CLT} the density of
\[
  \sum_{i=k+1}^{\varphi(n)} a_{i}^{(\varphi(n))}X_{i}
\]
is uniformly bounded. Hence by Lemma \ref{Sum has density},  $S_{\varphi(n)}$ has uniformly bounded density. This contradicts the initial assumption.
\smallskip
\subsubsection{\textbf{Case 3: At least \texorpdfstring{$5$}{5} bounded away from zero coefficients}}
\label{Case3}
\smallskip

There exist $i_1,\dots, i_5$ and \\ $\delta_1,\dots,\delta_5 > 0$ such that for $j\in\{1,\dots,5\}$, $|a_{i_j}^{(n)}| \ge \delta_j$ for infinitely many $n$. Without loss of generality, we can assume that $i_j = j$ and the inequalities hold for all $n$. Then, by Lemma \ref{Bound sum of five}, the density of $a_1^{(n)}X_1 + \dots + a_5^{(n)}X_5$ is bounded by $\varphi(\delta_1, \dots, \delta_5)$. Hence, by Lemma \ref{Sum has density}, $S_n$ has density bounded by $\varphi(\delta_1, \dots, \delta_5)$ for all $n$. This leads to a contradiction, and completes the proof.
\qed

\subsection{\textbf{Proof of Theorem \ref{mult-dim-deloc-thm}}}


The approach is similar to the proof of Theorem \ref{thm3}. Assume that contrary to the assertion of Theorem \ref{mult-dim-deloc-thm}, there exists a sequence of $S_n = \sum_{k=1}^{n} a_k^{(n)} X_k$ whose density functions $f_n$ are unbounded. By Lemma \ref{Bound affine} we may assume that $\mu_k = \E(X_k)=0$ for all $k$.

In the following we denote by $D_{\delta}$ the set of all symmetric, positive-definite $d\times d$ matrices with determinant at least $\delta$. It is a closed and convex set (see proof of Theorem 1 in \cite{GS}).

Let
\[
  C_n = \sum_{k=1}^{n}|a_k^{(n)}|^2\Gamma_k
\]
be the covariance matrix of $S_n$.

By convexity of $D_{\delta}$ we have $C_n\in D_{\delta}$ for every $n$. Moreover, all entries of $C_n$ are bounded by $L$. By the compactness argument we can assume without loss of generality that:
\[
  C_n \rightarrow C \in D_{\delta}.
\]
In particular, the matrix $C$ is  positive-definite.
\smallskip

\subsubsection{\textbf{Uniform decay of the coefficients}}\
\label{Case I}
\smallskip

Let's assume that 
\begin{equation}
  \label{vanish2}
\max_{1\le k\le n}{|a_k^{(n)}|} \rightarrow 0 \mbox{ as } n \to \infty.
\end{equation}
Then, by Theorem \ref{Local CLT} (see Appendix \ref{AppendixC})
\[
  \| f_n - g_C \|_{\infty} \rightarrow 0.
\]
Therefore, we have $\|f_n\|_{\infty} \le \frac{1}{2} + \|g_C\|_{\infty}$ for $n$ sufficiently large which leads to a contradiction.

\smallskip

\subsubsection{\textbf{At least \texorpdfstring{$1$}{1} bounded away from 0 coefficient}}
  \label{Case II}

If condition (\ref{vanish2}) is not true, then there exist $\gamma >0$, and sequences $\varphi(n)$ and $i(n)$ such that

\[
  |a_{i(n)}^{(\varphi(n))}| \ge \gamma
\]

Without loss of generality, we can assume that $i(n) = 1$ for every $n$. It follows from Lemma \ref{Sum has density} and \ref{Bound affine} that for every $n$ the density of $S_{\varphi(n)}$ is bounded by $\gamma^{-d}K$ which leads to a contradiction.



\qed


\section{\textbf{Concluding Remarks}}
\label{Section Concluding Remarks}

The main difference between our work and known results for LTI discrete channels is in the following:
\smallskip

\begin{itemize}
\item We argue in Section \ref{Section2} that it is reasonable, especially, in the scenario of wireless communication in urban environment, to assume
that the channel operator $H_N$ satisfies the PI assumption.
We have no assumptions on the power profile of the channel operator, i.e., no assumptions on the vector of $(\alpha_m)$'s.
\vspace{\baselineskip}
\item  The exact variational formula for the capacity is well known in the literature \cite{AC}, \cite{T}, \cite{TUV}. See also Formula (\ref{information}) in Section \ref{section-proof-main-thm}, and Formula 1.4 in \cite{TUV}. We provide a positive lower bound on the capacity under minor assumptions on the channel operators $H_N$.
\vspace{\baselineskip}
\item We relate the information theoretical problem at the hand to well established Delocalisation problem in Probability, and provide a new resolution of the latter through Lindeberg-Feller Local Central Limit Theorem \ref{Local CLT}. We also establish a completely new multi-dimensional delocalisation result, see Theorem \ref{mult-dim-deloc-thm}, which might have further applications in Information Theory.
\end{itemize}
\appendices
\section{Proof of Lemmata \ref{Sum has density} and \ref{Bound sum of five}}
\label{Appendix}

\begin{myproof}[Lemma \ref{Sum has density}]\

\begin{itemize}
\item \textbf{Absolute continuity of $aX+Y$.} First we need to show that $\Prob_{aX+Y}$ is absolutely continuous. Let $f$ denote the density of $Y$ and let $A$ be a Borel set with Lebesgue measure 0. Then by independence we have:
 \begin{align*}
  \Prob_{aX+Y}(A) & = \int_{\R^d}\Prob(Y\in A-x)d\Prob_{aX}(x) \\
          & = \int_{\R^d}\int_{A-x}f(y)dy d\Prob_{aX}(x) \\
          & = 0
 \end{align*}
In the last transition we used the fact that the Lebesgue measure is invariant by translation, so $A-x$ has Lebesgue measure $0$ for all $x \in \R^d$.
\item \textbf{Bound on the density of $aX+Y$}. Again, using the independence of $X$ and $Y$ and Fubini theorem, the following holds for any Borel set $A$:
 \begin{align*}
  \Prob_{aX+Y}(A) & = \int_{\R^d}\Prob(Y\in A-x)d\Prob_{aX}(x) \\
          & = \int_{\R^d}\int_{A-x}f(y)dy d\Prob_{aX}(x) \\
          & = \int_{\R^d}\int_{A}f(u-x)du d\Prob_{aX}(x) \\
          & = \int_{A}\int_{\R^d}f(u-x)d\Prob_{aX}(x)du \\
 \end{align*}
By uniqueness of the Radon-Nikodym derivative for almost every $u \in \R^d$, the density of the sum is defined by
 \begin{align*}
   g(u) & := \int_{\R^d}f(u-x)d\Prob_{aX}(x) \\
        & = \E(f(u-aX))
 \end{align*}

 Hence we have $ \| g \|_{\infty} \le \E(M) = M.$
 \end{itemize}

\qed
\end{myproof}

To prove Lemma \ref{Bound sum of five} we will need some estimates on Bessel function of the first kind. Denote by $\mu_r$ the uniform measure on the circle centered at $0$ of radius $r$, and by $\widehat{\mu_r}$ its Fourier transform defined for all $x\in\R^2$ as
\[
 \widehat{\mu_r}(x) = \int_{r S^1} e^{-i \langle y,x \rangle}d\mu_r(y)
\]

\begin{lemma}
 \label{Fourier transform of the uniform measure}
 $\widehat{\mu_r}$ is radial and for all $x\in\R^2$:
\[
 \widehat{\mu_r}(x) = \dfrac{1}{2\pi}\int_{0}^{2\pi}e^{-i r |x| cos(\theta)}d\theta
\]
The latter is also known as $J_0(r|x|)$ where $J_0$ is a Bessel function of the first kind.
\end{lemma}

\begin{myproof}[Lemma \ref{Fourier transform of the uniform measure}]
 It follows directly by passing to polar coordinates:
 \begin{align*}
  \widehat{\mu_r}(x) & = \dfrac{1}{2\pi}\int_{0}^{2\pi}e^{-i r |x| cos(\theta - Arg(x))}d\theta \\
         & = \dfrac{1}{2\pi}\int_0^{2\pi}e^{-i r|x|cos(\theta)}d\theta,
 \end{align*}
where the last transition is made by rotation invariance of the uniform measure on the unit circle.
 \qed
\end{myproof}

\begin{lemma}
 \label{Density and inverse Fourier}
 Let $X$ be an $\R^d-$valued random variable and let $\varphi$ denote its Fourier transform defined as $\varphi(y) := \E(e^{-i\langle y, X \rangle})$.
 If $\varphi \in L^1(\R^d)$ then $X$ has a density, bounded by $(2\pi)^{-d}\|\varphi\|_{L^1}$.
\end{lemma}
\begin{myproof}[Lemma \ref{Density and inverse Fourier}]
This is a standard fact about $\bR^d$-valued random variables.
\qed
\end{myproof}
%

\begin{lemma}
 \label{Bound Bessel}
 There exists $K$ such that for all $r > 0,\; |J_0(r)| \le \min\left(1, \frac{K}{\sqrt{r}}\right)$.
\end{lemma}

\begin{myproof}[Lemma \ref{Bound Bessel}]
 \begin{align*}
  J_0(r) & = \dfrac{1}{2\pi}\int_{-\pi/2}^{\pi/2}e^{-i r cos(\theta)}d\theta \\
      & + \dfrac{1}{2\pi}\int_{\pi/2}^{3\pi/2}e^{-i r cos(\theta)}d\theta \\
      & = \dfrac{1}{2\pi}(I_1 + I_2)
 \end{align*}
 By the Principle of Stationary Phase the following estimate holds as $r\to \infty$,
 \[
  I_k = \sqrt{\dfrac{2\pi}{r}}cos\left(r - \frac{\pi}{4}\right) + O\left(\frac{1}{r}\right)
 \]
 Hence $J_0(r) = \sqrt{\dfrac{2}{\pi r}}cos\left(r - \frac{\pi}{4}\right) + O\left(\frac{1}{r}\right)$. Therefore, $\sqrt{r}|J_0(r)|$ is bounded. We can take $K := \sup_{r>0}\sqrt{r}|J_0(r)|$.
 \qed
\end{myproof}

\begin{lemma}
 \label{Bessel L5}
$\widehat{\mu_a}\in L^5(\R^2)$ for any $a > 0$.
\end{lemma}

\begin{myproof}[Lemma \ref{Bessel L5}]
 $\widehat{\mu_a}$ is radial, continuous on $\R^2$ and by Lemma \ref{Bound Bessel}, $|\widehat{\mu_a}(x)|^5 = O\left(\dfrac{1}{|x|^{5/2}}\right)$
 as $|x| \rightarrow \infty$, which is integrable on $\R^2$. Therefore, so is $| \widehat{\mu_a} |^5$.
 \qed
\end{myproof}

We are now able to prove Lemma \ref{Bound sum of five}:

\begin{myproof}[Lemma \ref{Bound sum of five}]
 Let $\mu_a$ denote the probability measure of $aX$. By independence, the probability measure $\mu$ of the sum $a_1X_1 + \ldots + a_5X_5$ is the convolution $\mu_{a_1}\ast\ldots\ast\mu_{a_5}$.
 Then, $\widehat{\mu} = \widehat{\mu_{a_1}}\ldots\widehat{\mu_{a_5}}$. It follows from Lemmata \ref{Fourier transform of the uniform measure} and \ref{Bound Bessel} that
 \begin{align*}
  | \widehat{\mu}(x) | & = \left| J_0(|a_1x|)\ldots J_0(|a_5x|) \right| \\
           & \le  \prod_{k=1}^5 min\left(1, \dfrac{K}{\sqrt{a_k|x|}}\right) \\
           & \le  \prod_{k=1}^5 min\left(1, \dfrac{K}{\sqrt{b_k|x|}}\right) \\
 \end{align*}

 Hence,
 \[
  \|\widehat{\mu}\|_{L^1} \le \left\| \prod_{k=1}^5 min\left(1, \dfrac{K}{\sqrt{b_k|x|}}\right)\right\|_{L^1}
 \]

 Therefore, the result follows from Lemma \ref{Density and inverse Fourier}. \qed
\end{myproof}

\section{Proof of Theorem 
  \ref{Local CLT} 
}
\label{AppendixB}

In order to prove Theorem \ref{Local CLT}, it will suffice to prove that $(S_n)$ converges in distribution towards the Gaussian. This will hold true using an $\R^d$ version of Lindeberg-Feller Theorem that we recall hereafter.
We will denote by $\cvdis$ the convergence in distribution.

\begin{thm}[Lindeberg-Feller CLT]\label{Real LF}
  For each $n$ let $X_{n,m}$, $1\le m\le n$ be independent real-valued random variables with $\E[X_{n,m}]=0$. If

  \begin{enumerate}
  \item $\displaystyle \sum_{m=1}^n \E[X_{n,m}^2] \rightarrow \sigma^2$ as $n \rightarrow \infty$
  \item For each $\eps >0$,
    \begin{align*}
      \displaystyle \sum_{m=1}^n \E\left[\left|X_{n,m}\right|^2\indic_{\{|X_{n,m}|>\eps\}}\right] \rightarrow 0,\\
    \end{align*}
  \end{enumerate}

\noindent then $S_n := X_{n,1} + \dots + X_{n,n}\cvdis \Norm(0, \sigma^2)$.
\end{thm}
\begin{myproof}[Theorem \ref{Real LF}]
The proof can be found in Section 2.4 of \cite{Durrett}. \qed
\end{myproof}

Using Cram\'er-Wold's characterisation of convergence in distribution in $\R^d$, we can easily generalise Theorem
\ref{Real LF} to random vectors.

\begin{thm}[Cram\'er-Wold]\label{Cramér-Wold}
  Let $(X_n)$ be a sequence of random vectors in $\R^d$. We have $X_n \cvdis X_{\infty}$ if and only if $\langle\theta, X_n\rangle \ \cvdis \ \langle\theta, X_{\infty}\rangle$ for all $\theta\in \R^d$.
\end{thm}

\begin{thm}[$\R^d$-valued Lindeberg-Feller CLT]\label{Multivariate LF}
For each $n$ let $X_{n,m}$, $1\le m\le n,$ be independent $\R^d$-valued random vectors with $\E[X_{n,m}]=0$. Let $C_{n,m} := \E[X_{n,m} X_{n,m}^t]$ be the covariance matrix of $X_{n,m}$.

If we have

  \begin{enumerate}[label=\textbf{\roman*}), align=left]
  \item There exists $C$ a symmetric, positive-definite, $d\times d$ matrix such that :
    \[
      \displaystyle \sum_{m=1}^n C_{n,m} \rightarrow C
    \]
  \item For any $\eps>0$, for any $\theta\in\R^d$,
    \[
      \sum_{m=1}^n \E\left[|\langle \theta, X_{n,m} \rangle|^2\indic_{\{|\langle\theta, X_{n,m}\rangle|>\eps\}}\right] \rightarrow 0
    \]
  \end{enumerate}
then $S_{n} := X_{n,1} + \ldots + X_{n,n}$ converges in distribution towards a Gaussian with $0$ mean and covariance matrix $C$.

\end{thm}

Theorem \ref{Local CLT} is now a particular case of Theorem $1$ of \cite{Sherv}.

\section{} 
\label{AppendixC}


In what follows, we will check that all the requirements of Theorem \ref{Local CLT} are fulfilled.
\subsection{In the proof of Theorem \ref{thm3}:}

For the uniform distribution on the unit circle, the real and imaginary parts are uncorrelated. Moreover, their first moment is $0$ and their variance are equal. In what follows, let us denote by  $X_{n,m} = a_m^{(n)} X_m$, where $X_n$'s are independent uniformly distributed random variables on the unit circle in $\bC$. 

For each $m,$ and $n$ we have $C_{n,m} = \E(X_{n,m} X_{n,m}^t) = \dfrac{1}{2}\left|a_m^{(n)}\right|^2 I_2$, and therefore $\sigma^2_{n,m} = tr(C_{n,m}) = \left|a_m^{(n)}\right|^2$.
Hence,
\begin{align}
  \label{equality}
  \sum_{m=1}^nC_{n,m} = \dfrac{1}{2}I_2.
\end{align}
If we have
\[
  \max_{1\le m \le n} \left|a_m^{(n)}\right| \rightarrow 0,\mbox{ as } n \to \infty,
\]
then for every $\eps>0$ the event
\[
  \left\{\left|a_m^{(n)}X_m\right| > \eps\right\}
\]
is empty starting from some $n$. Then, Lindeberg-Feller's condition \ref{Feller} in Theorem \ref{Local CLT} holds true.

%
Moreover, by the results of Appendix \ref{Appendix}, the assumption on the Fourier transforms is satisfied for $p=5$, so the requirements of Theorem \ref{Local CLT} hold true.

\subsection{In the proof of Theorem \ref{mult-dim-deloc-thm}:}

Let $\theta\in\R^d$ and $\eps >0$. Then we have the following:

\begin{align*}
  & \; | \langle \theta, a_k^{(n)}X_k \rangle| ^2 \indic_{\{| \langle \theta, a_k^{(n)}X_k \rangle| > \eps\}} \\
  = & \; |a_k^{(n)}|^2 | \langle \theta, X_k \rangle| ^2 \indic_{\left\{\left|\dfrac{a_k^{(n)}\langle \theta, X_k \rangle}{\eps}\right|^{\eta} > 1\right\}} \\
  \le & \; \frac{1}{\eps^{\eta}}\left|a_k^{(n)}\right|^{2+\eta} \left| \langle \theta, X_k \rangle\right| ^{2+\eta}.
\end{align*}
Denote by $\gamma_n = \max_{1 \le k \le n} |a_{k}^{(n)}|$ and assume that $\gamma_n \to 0$ as $n \to \infty$. Then, by taking the expectation and using Cauchy-Schwarz inequality we have:

\begin{align*}
  \sum_{k=1}^n& \; \E[| \langle \theta, a_k^{(n)}X_k \rangle| ^2 \indic_{| \langle \theta, a_k^{(n)}X_k \rangle| > \eps}] \\
              & \; \le \frac{1}{\eps^\eta}     \sum_{k=1}^n \left|a_k^{(n)}\right|^{2+\eta} \E[| \langle \theta, X_k \rangle| ^{2+\eta}] \\
              & \; \le \frac{\|\theta\|^{2+\eta}}{\eps^\eta}\gamma_n^\eta\sum_{k=1}^n \left|a_k^{(n)}\right|^2\E\left[\|X_k\|^{2+\eta} \right]\\
              & \; \le \frac{M\|\theta\|^{2+\eta}}{\eps^\eta}\gamma_n^\eta \rightarrow 0\quad \text{as }n\rightarrow \infty
\end{align*}
Recall, we derived in the Proof of Theorem \ref{mult-dim-deloc-thm} that the covariance matrices of the sums $S_n$ converge to a positive-definite matrix $C$.
Therefore, all the requirements of Theorem \ref{Local CLT} are satisfied.  

%

\section*{Acknowledgment}

The authors would like to thank Ben Goldys, Georg Gottwald, Uri Keich, Mark Rudelson and Ofer Zeitouni for sharing their ideas on the various subjects related to the work. The first named author would like to thank the hospitality of the University of Sydney where this work has been carried out.

\begin{IEEEbiographynophoto}{Maxime Bombar}
is a 3rd year student at \'Ecole Normale Sup\'erieure Paris-Saclay, France, studying mathematics and computer science.
\end{IEEEbiographynophoto}
\begin{IEEEbiographynophoto}{Alexander Fish}
obtained a Ph.D. in mathematics in 2006 from Hebrew University,
Jerusalem, Israel. He conducts research in ergodic theory, and wireless
communication. He had post-doc positions in Ohio State University, Mathematical
Sciences Research Institute in Berkeley, and University of Wisconsin-Madison. From July 2012 he is a faculty member at the University of Sydney, Australia, in the School of Mathematics and Statistics.
\end{IEEEbiographynophoto}
\vfill

\end{document}